# Title:

# Biomimetic Metamaterial-based Interface for Decoding Heterogeneous Mechanodermal Activity

# Authors:

Muzi Xu[1], Jiaqi Zhang[2], Chaoqun Dong[1], Zibo Zhang[1], Duanyang Li[1], Wentian Yi[1], Miaomiao Zou[3], Chenyu Tang[1], George G. Malliaras[1], and Luigi G. Occhipinti*[1]

# Affiliations:

[1]Electrical Engineering Division, Department of Engineering, University of Cambridge; Cambridge, CB3 0FA, UK.

[2]Department of Electrical and Electronic Engineering, University of Hong Kong; Pokfulam Road, Hong Kong SAR, 999077, China.

[3]Institute for Manufacturing, Department of Engineering, University of Cambridge; Cambridge, CB3 0FS, UK.

*Corresponding author. Email: lgo23@cam.ac.uk

**Abstract:** Human skin acts as a dynamic biomechanical interface that conveys critical physiological and behavioural information through spatiotemporally distributed deformations. Due to the limited capabilities of current sensing technologies, the spatiotemporal diversity of its mechanical cues has remained underutilised to date, preventing these mechanisms from being used to capture and decode the full spectrum of underlying physiological states. In this work, we define this heterogeneous set of mechanical signals as mechanodermal activity (MDA) and introduce the biomimetic metamaterial-based interface (BMMI), an engineered auxetic metamaterial substrate that reproduces the microrelief and mechanoreceptor architecture of natural skin. The BMMI allows selective capture of diverse MDA signals from adjacent skin regions with simultaneous signal amplification and noise suppression, and permits straightforward modulation to accommodate various scenarios. Combined with bespoke algorithms, the wireless BMMI device decodes MDA accurately and robustly for multimodal communication interfaces, unleashing applications in healthcare monitoring and human-machine interaction.

**Main Text:**

Human skin serves as a dynamic biomechanical interface, reflecting a wide spectrum of mechanical deformations corresponding to physiological changes within the body. These deformations, spatiotemporally distributed across the skin, encode a rich array of biological information that provides non-invasive access to key indicators of physiological states and behaviours[1–7]. Recent research has predominantly focused on capturing and analysing single classes of biomechanical deformations, such as tracking hand movements for rapid gesture recognition[6,8] or detecting subtle arterial pressure for cardiac function assessment[9,10]. However, these approaches overlook the heterogeneous and spatially distributed nature of mechanical signals across adjacent skin regions, limiting the possibility of decoding the full spectrum of physiological information that can be obtained from skin activity, hindering a comprehensive understanding and automatic decoding of the body's dynamics and body language. To address this fundamental aspect, we define the property of the human body that causes the spatiotemporal distribution of mechanical deformations across the skin as mechanodermal activity (MDA), a unifying framework that integrates heterogeneous mechanical signals and enables new paradigms in healthcare monitoring, disease diagnosis, and human–machine interaction.

While electrodermal activity (EDA) has been widely studied and correlated to electrical activity occurring in the bulk, as in the case of electromyography (EMG) used to decode muscle activity, MDA occurs dynamically on the skin surface, where deformation of adjacent regions of skin produces signals and information with distinct characteristics. Taking the neck region as an example, spatially distributed heterogeneous signals, including small vibrations from the carotid artery, large deformations from lateral muscle movement, and small vibrations from laryngeal activity, can be captured simultaneously, providing physiological information associated with non-verbal communication[11]. Existing strain sensors on flexible and conformable substrates have primarily emphasized improvements in overall sensitivity, extending their single-point signal detection sensitivity by employing advanced nanomaterials[12–14] or novel structures[15–18]. These strategies, however, fail to capture the human skin's natural ability to produce spatially selective responses and parallel multi-signal processing[19], to directly capture high-fidelity signals at the source while simultaneously amplifying small vibrations and suppressing noise from large deformations across the skin. To effectively decode these heterogeneous MDAs, next-generation interfaces need to feature spatially selective signal acquisition capabilities with high specificity for both small vibrations and large deformations.

**Biomimetic Metamaterial-based Interface (BMMI)**

Here, we introduce a biomimetic metamaterial-based interface (BMMI) designed to achieve amplification of small vibrations and noise suppression of large deformations at the hardware level. This dual functionality is obtained through a biomimetic design approach incorporating an auxetic metamaterial, with dual tuneable Gauge Factor (GF) and detection limit, producing heterogeneous responses to multiple MDAs in the same substrate (Figure 1).

The BMMI mimics human skin in both surface morphology and tactile functionality. Skin microrelief refers to the intricate microscopic topography of the skin surface[20], composed of a network of furrows and ridges that form plateaux with triangular shapes[21]. As illustrated in Figure 1A, this microrelief enhances the skin's stretchability, allowing it to accommodate large strains induced by body movements without mechanical failure or tearing. Furthermore, the microrelief induces localized strain concentrations under compression, amplifying sensitivity to mechanical stimuli while attenuating responses in adjacent regions[21,22]. Inspired by this natural design, we implemented an auxetic metamaterial featuring star-shaped perforations as the structural backbone and applied strain engineering to create a heterogeneous functional substrate by modulating the Young's modulus (Figure 1B). Upon stretching, this architecture facilitates strain redistribution across the substrate. The star-shaped perforations filled with low-modulus materials promote localized strain concentration, while the high-modulus triangular backbone enables strain relaxation. This auxetic design also ensures excellent conformability to the skin's non-zero Gaussian surface.

Building on its biomimetic surface morphology, the BMMI substrate integrates regions of both strain concentration and relaxation. As shown in Figure 1C, the BMMI leverages the distinct properties of these two regions to replicate the tactile capabilities of human skin, achieving targeted sensing of both small vibrations, through BMMI-S, and large deformations, through BMMI-L, which correspond to the functions of Pacinian Corpuscles and Ruffini Endings, respectively. This design is directly inspired from the multimodal sensory system of human skin, which comprises low-threshold mechanoreceptors (LTMRs) with unique sensitivities and adaptive properties, enabling humans to perceive and respond to mechanical stimuli from the physical environment with high specificity[19,23]. Pacinian corpuscles, acting as fast-adapting receptors, are responsible for measuring vibrations transmitted through tissues, whereas Ruffini endings, as slowly-adapting receptors, primarily respond to skin stretching as result of tensile strain applied to skin[24,25]. Accordingly, BMMI-S, located in the low-modulus regions, exhibits a high gauge factor, allowing precise detection of subtle skin vibrations. In contrast, BMMI-L, situated in the high-modulus regions, possesses a low gauge factor, ensuring accurate sensing of large deformations. By seamlessly integrating heterogeneous mechanical responses within a single substrate, this biomimetic design creates a simple, efficient, and tuneable interface that selectively amplifies small vibrational signals while inherently suppressing excessive mechanical noise from large deformations, hence offering a unique and promising strategy for versatile and efficient decoding of diverse MDAs.

In Figure 1D, the exploded view depicts the structure and material composition of the fabricated BMMI, which consists of three main components: isolation frames, a sensing layer, and a functional substrate (FS1-3). FS-1, serving as the structural backbone, is designed as an auxetic metamaterial with star-shaped perforations and exhibits a high modulus optimized for responding to large deformations. In contrast, FS-2, designed for detecting small vibrations, is formed by filling the star-shaped perforations of FS-1 with a low-modulus material, which results in a continuous, integrated functional layer. FS-3, with a tissue-like modulus, serves the purpose of a bottom layer optimized for direct conformal contact with the skin to facilitate strain transduction. Finally, the graphene nanoplatelet-based sensing layer is deposited onto the surfaces of FS-1 and FS-2, forming the BMMI-L and BMMI-S sensors, respectively. A detailed fabrication procedure and cross-sectional schematic of the complete BMMI architecture are provided in Supplementary Fig.1 and Fig.2.

Using the fabricated BMMI integrated with a wireless electronics module, we demonstrated its effectiveness in multimodal communication analysis applications (Figure 1E). In the area of wearable electronics, conventional communication interfaces primarily focus on recognizing verbal information collected from the throat or mouth regions[2,26,27]. While these systems perform well in word identification and sentence decoding, they often neglect the non-verbal cues that are essential for a natural and effective human interaction. To overcome the limitations of these single-modality systems and enable a more comprehensive understanding of communicative intent, the BMMI was applied to the neck region to selectively capture spatiotemporally distributed, heterogeneous MDAs occurring in this region. In addition to verbal content, the BMMI simultaneously captured a wide array of non-verbal signals, including physiological data (pulse and respiration), auditory features (volume and speech rate), and postural cues linked to muscle activity (Supplementary Video 3). These signals, typically acquired using multiple sensors placed at various body sites, were now collected by a single device. The four-channel BMMI system wirelessly transmitted these heterogeneous signals, which were further decoded using a bespoke deep learning model to extract underlying cognitive and affective states such as attention, attitude, and emotion. As a result, the system achieved high-fidelity multimodal signal acquisition and enabled accurate recognition of these states with up to 95% accuracy, supporting more profound understanding and more effective communication.

**Modulation Mechanism of Metamaterials-based Substrate**

Mechanical metamaterials are artificially designed materials with engineered architectures that demonstrate properties surpassing those of conventional materials[28–30]. To break the existing fetters, these high-performance metamaterials have been widely applied in the field of flexible electronics for enhancing and innovating functionalities[31–33]. Among them, auxetic metamaterials, exhibiting a negative Poisson's ratio, offer superior biaxial stretchability, improved conformability to curved surfaces, and greater tunability compared to traditional elastomers[33–35]. Building on these advantages, an auxetic metamaterial with star-shaped perforations is utilized in this work to modulate the strain distribution in BMMI. Unlike other auxetic structures, such as re-entrant or chiral designs, the rotating triangle geometry provides sufficient space for both the high-modulus metamaterial backbone and perforations filled with low-modulus material to accommodate the sensing materials. As a result, such a biomimetic metamaterial-based substrate, integrated with strain engineering[36], achieves strain concentration in low-modulus regions and strain relaxation in high-modulus areas, enabling the selective detection of heterogeneous biophysical signals.

Figure 2A describes the mechanism of auxetic behaviour with rotating rigid triangles. Under tensile strain, these triangles begin to rotate, resulting in an increased angle between them, which forms a more open structure and achieves a negative Poisson's ratio. Leveraging this mechanism, we innovatively replaced the rigid triangles with high-modulus stretchable PDMS and filled the star-shaped perforations with low-modulus PDMS. This dual-modulus design not only retains the auxetic metamaterial's capability to modulate the Poisson's ratio but also converts the entire functional substrate into a flexible, stretchable interface exhibiting heterogeneous mechanical responses. As shown in Figure 2B, the cross-sectional view demonstrates that the high-modulus FS-1 and low-modulus FS-2 together form a continuous, integrated functional layer comprising multiple metamaterial units with tuneable geometric parameters. Specifically, as the r parameter, defined as the ratio of side lengths a to b, approaches the value of 1, the auxetic behaviour of the metamaterial is enhanced[37]. To maximize the auxetic performance of the functional substrate, r is therefore set to 1, corresponding to an equilateral triangular geometry. Previous studies have demonstrated that non-crystalline systems can exhibit auxetic behaviour comparable to that of rigid triangle-based designs, which is best achieved at lower values of the hinging angle ($\theta$) and the spacing (g) between adjacent star-shaped regions[37]. Based on this insight, the dimensions of our metamaterial units were defined as shown in Supplementary Fig.3, where $\theta$ is 30° and g is 1.5mm.

To investigate the tuneable strain distribution and auxetic behaviour of the BMMI functional substrate, both Finite Element Analysis (FEA) and tensile experiments were carried out. The results show that the heterogeneous strain distribution and the corresponding Poisson's ratio can be effectively modulated by tailoring the modulus ratio between FS-1 (high modulus) and FS-2 (low modulus), in agreement with our design analysis. Firstly, tensile simulations were performed on FS-1 alone to demonstrate the origin of its auxetic behaviour, showing that the selected unit geometry undergoes longitudinal expansion under various lateral tensile strains, as presented in Supplementary Fig.4. Secondly, functional substrates with five different modulus ratios between FS-1 and FS-2 (1:1, 6:1, 15:1, 52:1, and 144:1) were simulated under 5% strain to evaluate their strain modulation performance in Figure 2C, which is directly related to the heterogeneous biosignal responses of the BMMI. With increasing modulus ratio, the heterogeneity of the strain distribution is amplified, resulting in smaller values of strain in the "low-strain" regions (FS-1) and larger strains in the "high-strain" regions (FS-2), which allow for tuning the sensors' GF in both areas.

Furthermore, the Poisson's ratios of the five functional substrates, fabricated with varying polydimethilsiloxane (PDMS) modulus ratios, were characterized using a tensile tester, as shown in Figure 2D. The progressive decrease in Poisson's ratio (from 0.468 to –0.051) with increasing modulus ratio underscores the effective modulation enabled by the metamaterial backbone design, while simultaneously enhancing the conformability of the BMMI to the skin compared to a flat substrate (Supplementary Fig.5). In addition, all five functional substrates maintain stable mechanical performance (Supplementary Fig.6), attributed to the strong interfacial bonding between the constituent materials, as detailed in the Supplementary Text. Finally, the photographs in Figure 2E clearly illustrate the heterogeneous deformation of the BMMI, with smaller deformation in FS-1 and larger stretching in FS-2, under tensile strains ranging from 5% to 20%.

**Characterization of the BMMI**

Integrated with graphene nanoplatelets (prepared by High-Pressure Homogenization (HPH) method characterized in Supplementary Fig.7 and Fig.8) on both FS-1 and FS-2, the heterogeneous strain distribution of the BMMI is reflected in the distinct crack densities of the graphene layers. Supplementary Fig.9 presents the photograph of the fabricated BMMI, where FS-1 combined with the sensing layer forms the BMMI-S sensor, and FS-2 combined with the sensing layer forms the BMMI-L sensor. The SEM images in Figure 3A, scanned from the surfaces of the BMMI-S sensor (low modulus, sensitive to small vibrations) and the BMMI-L sensor (high modulus, responsive to large deformations) under 5% tensile strain, clearly reveal the differentiated crack formation resulting from strain engineering. Evidently, the strain-concentrated BMMI-S sensor exhibits longer, wider, and denser cracks compared to the strain-relaxed BMMI-L sensor. To facilitate visualization and comparison of the crack morphology, a quantitative analysis was conducted (Supplementary Fig.10), further confirming that the crack density in the graphene layers governs the heterogeneous sensing responses of the BMMI.

In the previous subsection, the modulatory effect of the modulus ratio on the functional substrate was analysed. Here, we quantitatively characterize how the modulus ratio regulates the strain-sensing capability of BMMI. All sensing signals are presented as relative resistance changes, with PS, D2, D3, and D4 corresponding to BMMIs fabricated with modulus ratios of 1:1, 6:1, 15:1, and 52:1, respectively. Figure 3B shows the performance of the BMMI-S sensors, where the relative resistance change gradually grows with the rising modulus ratio under identical cyclic stretching, indicating that the increasingly localized strain in BMMI-S boosts its sensitivity to small vibrations. Conversely, Figure 3D presents the BMMI-L sensors' performance, where the relative resistance change steadily declines with higher modulus ratios, suggesting that the more relaxed strain in BMMI-L effectively reduces minor noise interference during large deformations. As a result, the two opposite trends of the Gauge Factor (GF) curves in Figure 3C summarize how the modulus ratio modulates the BMMI's performance. Compared to the plain sensor (PS), our designed BMMI achieves dual tuneable GFs, with the GF of BMMI-S increasing by up to approximately 12 times, and that of BMMI-L decreases by about 4 times, thereby realizing heterogeneous sensing for small vibrations and large deformations with intrinsic capabilities for both signal enhancement and noise suppression.

Taking advantage of its dual tunability, the BMMI can be customized to suit specific application scenarios. In this work, we selected D3 (modulus ratio 15:1) for the detection of diverse biomechanical signals on the neck, where the GFs of BMMI-S and BMMI-L reach 41 and 405, respectively. As demonstrated by the relative resistance changes in Figure 3E and Figure 3F, the BMMI achieves a tenfold difference in GF within a single functional substrate, with BMMI-S excelling in small-strain regions and BMMI-L performing well for large-strain regions. Both BMMI sensors demonstrate consistent and stable responses under cyclic stretching with varying strain amplitudes (Supplementary Fig.11 and Fig.12), as well as outstanding durability over 10,000 cycles (Supplementary Fig.13 and Fig.14). In addition, BMMI-S and BMMI-L display distinct detection limits, with BMMI-S capable of sensing strains as small as 0.01%, while BMMI-L detects strains down to 0.1% (Figure 3G).

**Detection of Heterogeneous MDA**

To assess the superior capabilities of the BMMI in capturing spatially heterogeneous signals on the skin surface under real-world conditions, two BMMI-S and two BMMI-L sensors were systematically distributed across four metamaterial units corresponding to the neck region for the simultaneous acquisition of multiple adjacent physiological signals. When the user wears the four-channel BMMI, a broad spectrum of mechanical deformations, including small vibrations generated by the carotid artery and throat region, and large deformations from the lateral neck muscles, can be selectively captured by BMMI-S and BMMI-L, respectively.

Specifically, the BMMI-S sensor reliably detects subtle pulse signals with regular intervals in the carotid artery region (Figure 4A). Additionally, owing to the unique anatomical characteristics of the neck, the BMMI-S simultaneously captures the respiratory cyclic waveform superimposed on the pulse signal (Figure 4B), enabling concurrent assessment of both cardiovascular and respiratory activities. Meanwhile, the central BMMI-L sensors capture a variety of head movements, encompassing both static and dynamic motions. As shown in Figure 4C, static actions, such as turning the head left by 10° or 20°, and dynamic movements like nodding and shaking the head are distinctly differentiated and recorded. Notably, the BMMI-L sensor also resolves compound mechanical events. For instance, Figure 4D presents the composite signal generated by nodding or shaking the head while it is held at a 20° or 10° leftward rotation. Finally, another BMMI-S sensor placed over the laryngeal region precisely discerns variations in spoken words, speech rates, and loudness levels. Two representative words, "Electrical" and "Graphene," are evaluated under three different speaking rates and three loudness levels. As illustrated in Figure 4E and Figure 4F, although the same word maintains consistent waveform features across varying conditions, the signals exhibit distinct scaling in both temporal duration and amplitude, underscoring the remarkable sensitivity of the BMMI-S to subtle vocal vibrations.

The four-channel BMMI successfully acquired high-fidelity and high-quality multimodal biosignals from the neck region. This performance benefits from spatially targeted sensing, which amplifies signal magnitude and effectively suppresses unwanted noise, thereby establishing a robust foundation for reliable cognitive and affective analyses based on diverse physiological information. To more intuitively demonstrate the advantages of heterogeneous sensing, comparative experiments were conducted using plain sensors. In the context of arterial pulse detection, enhanced sensor sensitivity is critical for precise monitoring, which underpins effective physiological analysis[38]. When employing the BMMI-S sensor with a gauge factor as high as 405, the acquired pulse waveforms (Figure 4G) exhibit an idealized pattern with three gradually weakening positive peaks. These features enable detailed extraction of arterial pulse dynamics[38]. In contrast, the plain sensor produced signals with approximately half the amplitude and failed to preserve many critical waveform details. Moreover, the BMMI-L sensor demonstrated excellent resistance to motion artefacts in static conditions, as evidenced by a comparison with a plain sensor placed on the lateral neck during head rest (Figure 4H). These findings collectively highlight the exceptional capability of the BMMI system in the broad detection of heterogeneous MDA.

With the aim of reducing crosstalk between the BMMI-S and BMMI-L sensors, polyimide isolation frames were incorporated around the two BMMI-S sensors. The Finite element analysis (FEA) results shown in Supplementary Fig.15 indicate that when tensile strain is applied from both ends, the strain within the isolated region enclosed by the frames is significantly reduced. At the same time, high sensitivity is maintained inside the frame. This finding is consistent with the results presented in Figure 4I, where crosstalk from head rotation was substantially mitigated, effectively preserving the integrity of the pulse waveform.

**Multimodal Communication Analysis**

Humans rely on verbal and non-verbal communication to convey information comprehensively in daily life. Multimodal communication analysis plays a crucial role in understanding and enhancing human interaction by interpreting various forms of communication. Here, we used our wireless BMMI (Supplementary Fig.16) in conjunction with a novel deep learning model, CA-Net, to perform multimodal communication analysis for inferring internal cognitive-affective states, including emotion, attitudes, and attention.

Cognition and affect can be decoded from a range of high-fidelity physiological and behavioural signals, such as pulse, respiration, speech, and muscle movement[39–42]. Given the complex interplay among diverse biosignals that contribute to inherently multidimensional internal states, targeted multimodal perception is essential for robust and accurate recognition. Leveraging its unique ability to detect heterogeneous MDA, the four-channel BMMI can simultaneously capture multiple biophysical signals via BMMI-S and BMMI-L sensors placed at different regions of the neck, each exhibiting distinct strengths and characteristics (Figure 5A and Figure 5B). These valuable signals contain substantial spatiotemporal details hidden in physiological activities, which can be utilized for comprehensive analysis of the human cognitive and affective states.

Based on powerful data-driven deep learning techniques, we propose a novel model, CA-Net (Figure 5C), which incorporates a hybrid architecture encoder and three parallel output heads for precise classification of emotional, attitudinal, and attentional states. This approach, which fully considers the spatiotemporal diversity of the four-channel sequential signals—Channel 1 for detecting subtle carotid pulse and respiratory activity; Channels 2 and 3 for capturing static and dynamic head movements; and Channel 4 for recording laryngeal vibrations—adopts a hybrid backbone that integrates convolutional neural networks (CNNs)[43] and Transformers[44], thereby enabling cascaded feature extraction and cross-channel fusion.

Specifically, our CNN encoder features a quad-branch architecture, where each branch comprises three convolutional layers, each of them followed by a BatchNorm and a ReLU layers, enabling efficient local feature extraction from each channel through the strong inductive bias of convolution inspired by PESNet[45]. Additionally, the hierarchical outputs shorten the sequence length, thereby reducing computational overhead for the subsequent Transformer encoder. The four resulting feature sequences are then concatenated along the hidden dimension (D) and fed into the Transformer to establish long-range dependencies. For the Transformer stage, the model omits the original positional embedding process for each feature sequence to avoid the order of input channels on the consistency of prediction. Accordingly, the Transformer encoder is designed with two layers, each containing a multi-head self-attention head, which utilizes its global receptive field and consistent resolution to establish global interactions among all feature tokens from the four signal channels. Finally, an average pooling layer condensed the hidden dimension (D) to feed the fused features into parallel classification heads for specific tasks.

A total of 1200 sets of 4-channel data were collected from 5 subjects, encompassing three emotional states (happy, sad, calm), four attitudes (support, oppose, concern, skeptical), and five levels of attentional engagement (focused ×2, partially engaged, distracted ×2), with 100 sets for each category (Supplementary Figs.17-19). Of the total dataset, 70% was used for training the model, while 20% and 10% were allocated for validation and testing, respectively. As a result, our proposed CN-Net achieved classification accuracies of 95% for emotion, 100% for attitude, and 99% for attention recognition. The corresponding confusion matrices are displayed in Figure 5D-5F. To further validate the effectiveness of CA-Net in cognition and affect recognition, we conducted comparative experiments against conventional RNN- and pure Transformer-based models, as presented in Supplementary Table 1. CA-Net consistently outperformed these models in classification accuracy, showcasing its outstanding ability to capture spatiotemporal patterns across multimodal biosignals.

The key advantage and characteristics of multimodal integration were assessed through an ablation study in which individual channels were selectively used, or the number of input signals was reduced, to determine the contribution of the four-channel data to the overall classification performance. For example, in emotion recognition, using only the single-channel data from Channel 1 at the carotid artery, following conventional approaches, resulted in an accuracy of just 80%, significantly lower than the 95% achieved with the multimodal setup. Moreover, when heterogeneous MDA was excluded and only the small vibration signals from Channels 1 and 4 were used, the accuracy dropped to 91.67%, indicating that critical affect-related features were not fully captured. As shown in Figure 5G, similar ablation studies were conducted for the other two internal states, yielding consistent trends, with the corresponding confusion matrices presented in Supplementary Figs.20-22. These results demonstrate that although classification remains feasible with single-modality input, the absence of multimodal synergy leads to substantial degradation in performance. Therefore, by simultaneously acquiring spatially and temporally distributed heterogeneous MDAs using our BMMI system, we fully leveraged the synergistic effect of multimodal integration to decode and recognize complex internal states through the bespoke CA-Net model, achieving consistently higher classification accuracy across all categories.

**Conclusions**

By mimicking both the surface morphology and the tactile functionality of human skin, we have introduced a BMMI device which enables high-precision detection of spatiotemporally distributed heterogeneous MDA. Applied to a conformable artificial skin device, this capability allows for targeted capture of adjacent skin deformations associated with a range of physiological information of crucial importance for deeper understanding of human behaviour and non-verbal communication. Characterized by its unique auxetic metamaterial backbone designed for strain engineering, the BMMI goes beyond any existing concepts of flexible human-machine interface, and achieves hardware-level signal amplification and noise suppression on the same functional substrate, with application-specific tuneable performance. As a multimodal communication interface, based on its high-fidelity multichannel biophysical signals and the novel CA-Net model, the wireless BMMI system achieves unprecedented performance in the automatic analysis and classification of cognitive and affective states. The ablation studies conducted here provide new insights in the significance of the synergistic effect of heterogeneous biophysical signals associated with MDA. Building upon the technology platform introduced in this study to decode heterogeneous MDA, future efforts could extend the approach with further adjustments to the modulus ratio, exploring other geometric configurations and distributions of sensing materials, opening new avenues for health diagnosis, natural human-machine interaction, and beyond.

# References:

1. S. Lee, S. Franklin, F. A. Hassani, T. Yokota, M. O. G. Nayeem, Y. Wang, R. Leib, G. Cheng, D. W. Franklin, T. Someya, Nanomesh pressure sensor for monitoring finger manipulation without sensory interference. *Science* **370**, 966–970 (2020).

2. Q. Yang, W. Jin, Q. Zhang, Y. Wei, Z. Guo, X. Li, Y. Yang, Q. Luo, H. Tian, T.-L. Ren, Mixed-modality speech recognition and interaction using a wearable artificial throat. *Nat Mach Intell* **5**, 169–180 (2023).

3. O. A. Araromi, M. A. Graule, K. L. Dorsey, S. Castellanos, J. R. Foster, W.-H. Hsu, A. E. Passy, J. J. Vlassak, J. C. Weaver, C. J. Walsh, R. J. Wood, Ultra-sensitive and resilient compliant strain gauges for soft machines. *Nature* **587**, 219–224 (2020).

4. Q. Wang, H. Guan, C. Wang, P. Lei, H. Sheng, H. Bi, J. Hu, C. Guo, Y. Mao, J. Yuan, M. Shao, Z. Jin, J. Li, W. Lan, A wireless, self-powered smart insole for gait monitoring and recognition via nonlinear synergistic pressure sensing. *Science Advances* **11**, eadu1598 (2025).

5. Y. Kim, J. M. Suh, J. Shin, Y. Liu, H. Yeon, K. Qiao, H. S. Kum, C. Kim, H. E. Lee, C. Choi, H. Kim, D. Lee, J. Lee, J.-H. Kang, B.-I. Park, S. Kang, J. Kim, S. Kim, J. A. Perozek, K. Wang, Y. Park, K. Kishen, L. Kong, T. Palacios, J. Park, M.-C. Park, H. Kim, Y. S. Lee, K. Lee, S.-H. Bae, W. Kong, J. Han, J. Kim, Chip-less wireless electronic skins by remote epitaxial freestanding compound semiconductors. *Science* **377**, 859–864 (2022).

6. K. K. Kim, M. Kim, K. Pyun, J. Kim, J. Min, S. Koh, S. E. Root, J. Kim, B.-N. T. Nguyen, Y. Nishio, S. Han, J. Choi, C.-Y. Kim, J. B.-H. Tok, S. Jo, S. H. Ko, Z. Bao, A substrate-less nanomesh receptor with meta-learning for rapid hand task recognition. *Nature Electronics* **6**, 64–75 (2023).

7. C. Tang, W. Yi, M. Xu, Y. Jin, Z. Zhang, X. Chen, C. Liao, M. Kang, S. Gao, P. Smielewski, L. G. Occhipinti, A deep learning–enabled smart garment for accurate and versatile monitoring of sleep conditions in daily life. *Proceedings of the National Academy of Sciences* **122**, e2420498122 (2025).

8. Z. Zhou, K. Chen, X. Li, S. Zhang, Y. Wu, Y. Zhou, K. Meng, C. Sun, Q. He, W. Fan, E. Fan, Z. Lin, X. Tan, W. Deng, J. Yang, J. Chen, Sign-to-speech translation using machine-learning-assisted stretchable sensor arrays. *Nat Electron* **3**, 571–578 (2020).

9. M. Xu, J. Zhang, C. Dong, C. Tang, F. Hu, G. G. Malliaras, L. G. Occhipinti, Simultaneous Isotropic Omnidirectional Hypersensitive Strain Sensing and Deep Learning-Assisted Direction Recognition in a Biomimetic Stretchable Device. *Advanced Materials* **37**, 2420322 (2025).

10. S. Li, H. Wang, W. Ma, L. Qiu, K. Xia, Y. Zhang, H. Lu, M. Zhu, X. Liang, X.-E. Wu, H. Liang, Y. Zhang, Monitoring blood pressure and cardiac function without positioning via a deep learning–assisted strain sensor array. *Science Advances* **9**, eadh0615 (2023).

11. T. D. Royce, W. L. Bowcher, Eds., *New Directions in the Analysis of Multimodal Discourse* (L. Erlbaum Associates, Mahwah, N.J, 2007).

12. S. Gong, X. Zhang, X. A. Nguyen, Q. Shi, F. Lin, S. Chauhan, Z. Ge, W. Cheng, Hierarchically resistive skins as specific and multimetric on-throat wearable biosensors. *Nat. Nanotechnol.* **18**, 889–897 (2023).

13. D. H. Lee, T. Miyashita, Y. Xuan, K. Takei, Ultrasensitive and Stretchable Strain Sensors Based on Laser-Induced Graphene With ZnO Nanoparticles. *ACS Nano* **18**, 32255–32265 (2024).

14. S. Zhao, X. Meng, L. Liu, W. Bo, M. Xia, R. Zhang, D. Cao, J.-H. Ahn, Polypyrrole-coated copper nanowire-threaded silver nanoflowers for wearable strain sensors with high sensing performance. *Chemical Engineering Journal* **417**, 127966 (2021).

15. D. Kang, P. V. Pikhitsa, Y. W. Choi, C. Lee, S. S. Shin, L. Piao, B. Park, K.-Y. Suh, T. Kim, M. Choi, Ultrasensitive mechanical crack-based sensor inspired by the spider sensory system. *Nature* **516**, 222–226 (2014).

16. C. Tang, M. Xu, W. Yi, Z. Zhang, E. Occhipinti, C. Dong, D. Ravenscroft, S.-M. Jung, S. Lee, S. Gao, J. M. Kim, L. G. Occhipinti, Ultrasensitive textile strain sensors redefine wearable silent speech interfaces with high machine learning efficiency. *npj Flex Electron* **8**, 1–11 (2024).

17. X. Huang, L. Liu, Y. H. Lin, R. Feng, Y. Shen, Y. Chang, H. Zhao, High-stretchability and low-hysteresis strain sensors using origami-inspired 3D mesostructures. *Sci. Adv.* **9**, eadh9799 (2023).

18. L. Li, Y. Zheng, E. Liu, X. Zhao, S. Yu, J. Wang, X. Han, F. Xu, Y. Cao, C. Lu, H. Gao, Stretchable and ultrasensitive strain sensor based on a bilayer wrinkle-microcracking mechanism. *Chemical Engineering Journal* **437**, 135399 (2022).

19. A. Handler, D. D. Ginty, The mechanosensory neurons of touch and their mechanisms of activation. *Nat Rev Neurosci* **22**, 521–537 (2021).

20. T. Li, H. Qi, C. Zhao, Z. Li, W. Zhou, G. Li, H. Zhuo, W. Zhai, Robust skin-integrated conductive biogel for high-fidelity detection under mechanical stress. *Nat Commun* **16**, 88 (2025).

21. G. Limbert, E. Kuhl, On skin microrelief and the emergence of expression micro-wrinkles. *Soft Matter* **14**, 1292–1300 (2018).

22. W. Sun, B. Wang, T. Yang, R. Yin, F. Wang, H. Zhang, W. Zhang, Three-Dimensional Bioprinted Skin Microrelief and Its Role in Skin Aging. *Biomimetics* **9**, 366 (2024).

23. A. Zimmerman, L. Bai, D. D. Ginty, The gentle touch receptors of mammalian skin. *Science* **346**, 950–954 (2014).

24. R. S. Johansson, J. R. Flanagan, Coding and use of tactile signals from the fingertips in object manipulation tasks. *Nat Rev Neurosci* **10**, 345–359 (2009).

25. A. Chortos, J. Liu, Z. Bao, Pursuing prosthetic electronic skin. *Nature Mater* **15**, 937–950 (2016).

26. T. Kim, Y. Shin, K. Kang, K. Kim, G. Kim, Y. Byeon, H. Kim, Y. Gao, J. R. Lee, G. Son, T. Kim, Y. Jun, J. Kim, J. Lee, S. Um, Y. Kwon, B. G. Son, M. Cho, M. Sang, J. Shin, K. Kim, J. Suh, H. Choi, S. Hong, H. Cheng, H.-G. Kang, D. Hwang, K. J. Yu, Ultrathin crystalline-silicon-based strain gauges with deep learning algorithms for silent speech interfaces. *Nat Commun* **13**, 5815 (2022).

27. T. Liu, M. Zhang, Z. Li, H. Dou, W. Zhang, J. Yang, P. Wu, D. Li, X. Mu, Machine learning-assisted wearable sensing systems for speech recognition and interaction. *Nat Commun* **16**, 2363 (2025).

28. T. Frenzel, M. Kadic, M. Wegener, Three-dimensional mechanical metamaterials with a twist. *Science* **358**, 1072–1074 (2017).

29. J. B. Berger, H. N. G. Wadley, R. M. McMeeking, Mechanical metamaterials at the theoretical limit of isotropic elastic stiffness. *Nature* **543**, 533–537 (2017).

30. X. Fang, J. Wen, L. Cheng, D. Yu, H. Zhang, P. Gumbsch, Programmable gear-based mechanical metamaterials. *Nat. Mater.* **21**, 869–876 (2022).

31. Q. Zeng, X. Tian, D. T. Nguyen, C. Li, P. Chia, B. C. K. Tee, C. Wu, J. S. Ho, A digitally embroidered metamaterial biosensor for kinetic environments. *Nat Electron* **7**, 1025–1034 (2024).

32. D. Hwang, C. Lee, X. Yang, J. M. Pérez-González, J. Finnegan, B. Lee, E. J. Markvicka, R. Long, M. D. Bartlett, Metamaterial adhesives for programmable adhesion through reverse crack propagation. *Nat. Mater.* **22**, 1030–1038 (2023).

33. S. Jiang, X. Liu, J. Liu, D. Ye, Y. Duan, K. Li, Z. Yin, Y. Huang, Flexible Metamaterial Electronics. *Advanced Materials* **34**, 2200070 (2022).

34. K. Bertoldi, V. Vitelli, J. Christensen, M. van Hecke, Flexible mechanical metamaterials. *Nat Rev Mater* **2**, 1–11 (2017).

35. H. M. A. Kolken, A. A. Zadpoor, Auxetic mechanical metamaterials. *RSC Advances* **7**, 5111–5129 (2017).

36. Y. Jiang, Z. Liu, C. Wang, X. Chen, Heterogeneous Strain Distribution of Elastomer Substrates To Enhance the Sensitivity of Stretchable Strain Sensors. *Acc. Chem. Res.* **52**, 82–90 (2019).

37. J. N. Grima, R. Gatt, B. Ellul, E. Chetcuti, Auxetic behaviour in non-crystalline materials having star or triangular shaped perforations. *Journal of Non-Crystalline Solids* **356**, 1980–1987 (2010).

38. S. Min, J. An, J. H. Lee, J. H. Kim, D. J. Joe, S. H. Eom, C. D. Yoo, H.-S. Ahn, J.-Y. Hwang, S. Xu, J. A. Rogers, K. J. Lee, Wearable blood pressure sensors for cardiovascular monitoring and machine learning algorithms for blood pressure estimation. *Nat Rev Cardiol*, 1–20 (2025).

39. M. T. V. Yamuza, J. Bolea, M. Orini, P. Laguna, C. Orrite, M. Vallverdú, R. Bailón, Human Emotion Characterization by Heart Rate Variability Analysis Guided by Respiration. *IEEE Journal of Biomedical and Health Informatics* **23**, 2446–2454 (2019).

40. B. M. Appelhans, L. J. Luecken, Heart Rate Variability as an Index of Regulated Emotional Responding. *Review of General Psychology* **10**, 229–240 (2006).

41. K. Yang, C. Wang, Y. Gu, Z. Sarsenbayeva, B. Tag, T. Dingler, G. Wadley, J. Goncalves, Behavioral and Physiological Signals-Based Deep Multimodal Approach for Mobile Emotion Recognition. *IEEE Transactions on Affective Computing* **14**, 1082–1097 (2023).

42. Y. Hu, B. Chen, J. Lin, Y. Wang, Y. Wang, C. Mehlman, H. Lipson, Human-robot facial coexpression. *Science Robotics* **9**, eadi4724 (2024).

43. K. He, X. Zhang, S. Ren, J. Sun, "Deep residual learning for image recognition" in *Proceedings of the IEEE Conference on Computer Vision and Pattern Recognition* (2016; http://openaccess.thecvf.com/content_cvpr_2016/html/He_Deep_Residual_Learning_CVPR_2016_paper.html), pp. 770–778.

44. A. Vaswani, N. Shazeer, N. Parmar, J. Uszkoreit, L. Jones, A. N. Gomez, Ł. ukasz Kaiser, I. Polosukhin, "Attention is All you Need" in *Advances in Neural Information Processing Systems* (Curran Associates, Inc., 2017; https://proceedings.neurips.cc/paper/2017/hash/3f5ee243547dee91fbd053c1c4a845aa-Abstract.html)vol. 30.

45. J. Zhang, Y. Hu, X. Qi, T. Meng, L. Wang, H. Fu, M. Yang, J. Liu, "Polar Eyeball Shape Net for 3D Posterior Ocular Shape Representation" in *Medical Image Computing and Computer Assisted Intervention – MICCAI 2023*, H. Greenspan, A. Madabhushi, P. Mousavi, S. Salcudean, J. Duncan, T. Syeda-Mahmood, R. Taylor, Eds. (Springer Nature Switzerland, Cham, 2023; https://link.springer.com/10.1007/978-3-031-43987-2_18)vol. 14225 of *Lecture Notes in Computer Science*, pp. 180–190.

46. A. Sharfeddin, A. A. Volinsky, G. Mohan, N. D. Gallant, Comparison of the macroscale and microscale tests for measuring elastic properties of polydimethylsiloxane. *J of Applied Polymer Sci* **132**, app.42680 (2015).

47. I. D. Johnston, D. K. McCluskey, C. K. L. Tan, M. C. Tracey, Mechanical characterization of bulk Sylgard 184 for microfluidics and microengineering. *J. Micromech. Microeng.* **24**, 035017 (2014).

48. Y.-S. Yu, Y.-P. Zhao, Deformation of PDMS membrane and microcantilever by a water droplet: Comparison between Mooney–Rivlin and linear elastic constitutive models. *Journal of Colloid and Interface Science* **332**, 467–476 (2009).

49. O. Akogwu, D. Kwabi, S. Midturi, M. Eleruja, B. Babatope, W. O. Soboyejo, Large strain deformation and cracking of nano-scale gold films on PDMS substrate. *Materials Science and Engineering: B* **170**, 32–40 (2010).

50. A. E. Forte, P. Z. Hanakata, L. Jin, E. Zari, A. Zareei, M. C. Fernandes, L. Sumner, J. Alvarez, K. Bertoldi, Inverse Design of Inflatable Soft Membranes Through Machine Learning. *Adv Funct Materials* **32**, 2111610 (2022).

51. Y. Yu, D. Sanchez, N. Lu, Work of adhesion/separation between soft elastomers of different mixing ratios. *J. Mater. Res.* **30**, 2702–2712 (2015).

52. M. P. Wolf, G. B. Salieb-Beugelaar, P. Hunziker, PDMS with designer functionalities—Properties, modifications strategies, and applications. *Progress in Polymer Science* **83**, 97–134 (2018).

53. B. Ruben, M. Elisa, L. Leandro, M. Victor, G. Gloria, S. Marina, S. Mian K, R. Pandiyan, L. Nadhira, Oxygen plasma treatments of polydimethylsiloxane surfaces: effect of the atomic oxygen on capillary flow in the microchannels. *Micro & Nano Letters* **12**, 754–757 (2017).

54. A. Borók, K. Laboda, A. Bonyár, PDMS Bonding Technologies for Microfluidic Applications: A Review. *Biosensors* **11**, 292 (2021).

55. L. Francioso, C. De Pascali, R. Bartali, E. Morganti, L. Lorenzelli, P. Siciliano, N. Laidani, PDMS/Kapton Interface Plasma Treatment Effects on the Polymeric Package for a Wearable Thermoelectric Generator. *ACS Appl. Mater. Interfaces* **5**, 6586–6590 (2013).

**Acknowledgments:**

    **Funding:**

        British Council, Contract No. 45371261 (LGO)

UK Engineering and Physical Science Research Council (EPSRC) No. EP/K03099X/1 (LGO)

UK Engineering and Physical Science Research Council (EPSRC) No. EP/W024284/1 (LGO)

Haleon through the CAPE partnership contract, University of Cambridge No. G110480 (LGO)

**Author contributions:**

Conceptualization: M.X., L.G.O.;

Methodology: M.X., J.Z., C.D., L.G.O.;

Formal Analysis: M.X., J.Z.;

Investigation: M.X., J.Z., C.D., Z.Z., D.L., W.Y., M.Z., C.T.;

Software: M.X., J.Z.;

Visualization: M.X., J.Z.;

Validation: M.X., J.Z.;

Funding acquisition: L.G.O.;

Project administration: M.X., L.G.O.;

Supervision: M.X., L.G.O.;

Writing – original draft: M.X.;

Writing – review & editing: M.X., J.Z., G.G.M., L.G.O.

**Competing interests:** Authors declare that they have no competing interests.

**Data and materials availability:** All data are available in the main text and supplementary materials. Additional code is available from the corresponding author upon reasonable request.

**Supplementary Materials**

Materials and Methods

Supplementary Text

Figs. S1 to S22

Tables S1

References (*46*–*55*)

Movies S1 to S3

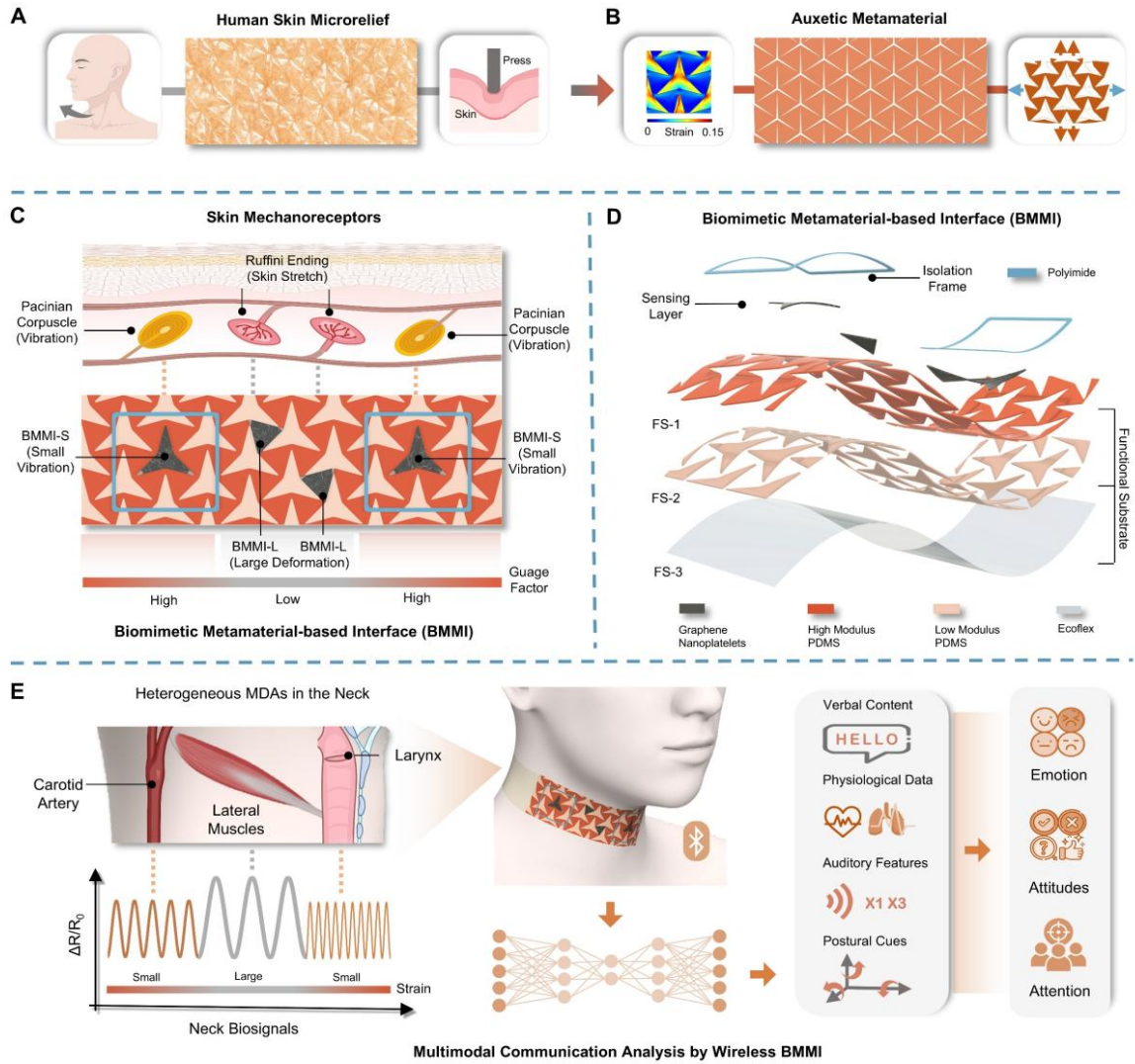

**Figure 1. Biomimetic Metamaterial-based Interface (BMMI).** (A) Photograph of the human skin microrelief, which accommodates large strains induced by body movements and generates localized strain concentrations under compression. (B) An auxetic metamaterial featuring star-shaped perforations, facilitating strain redistribution upon stretching. (C) The BMMI mimics the targeted tactile capabilities of human skin. BMMI-S and BMMI-L correspond to skin mechanoreceptors, namely Pacinian corpuscles and Ruffini endings, enabling the selective detection of small vibrations and large deformations, respectively. (D) The structure of the BMMI consists of a functional substrate (FS-1, FS-2, and FS-3), a sensing layer, and isolation frames. (E) Multimodal communication analysis using a novel deep learning-based model with four-channel heterogeneous data collected by a wireless BMMI attached to the neck region, enabling accurate recognition of cognitive and affective states. Some elements in Schematics (C) and (E) were created using BioRender.com.

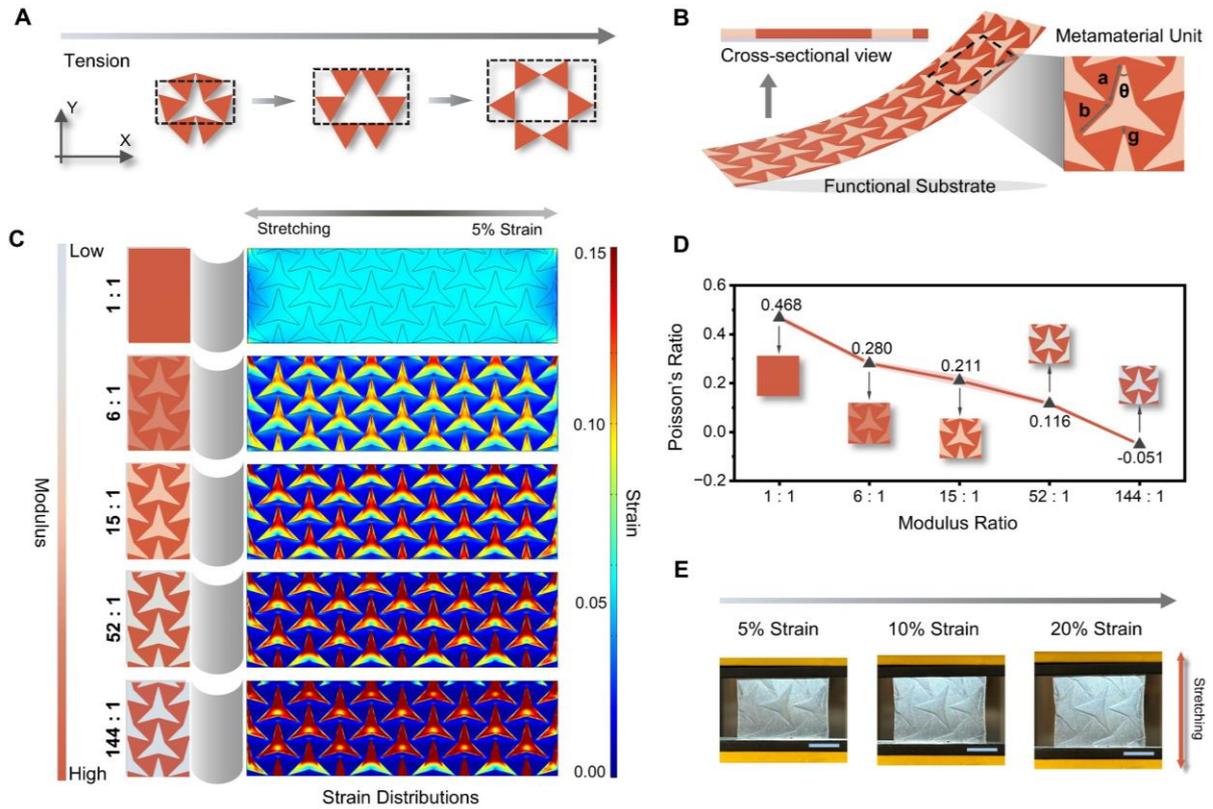

**Figure 2. Modulation mechanism of metamaterials-based substrate.** (A) The mechanism of auxetic behavior with rotating rigid triangles. Under tensile strain, these triangles rotate, forming a more open structure and achieving a negative Poisson's ratio. (B) The cross-sectional view and metamaterial unit of the BMMI's functional substrate. (C) Finite element analysis results of functional substrates with five different modulus ratios between FS-1 and FS-2 (1:1, 6:1, 15:1, 52:1, and 144:1) under 5% strain. (D) Poisson's ratios of the five functional substrates, fabricated with varying modulus ratios. (E) Photographs of the BMMI's functional substrate under different tensile tests. Scale bar 1cm.

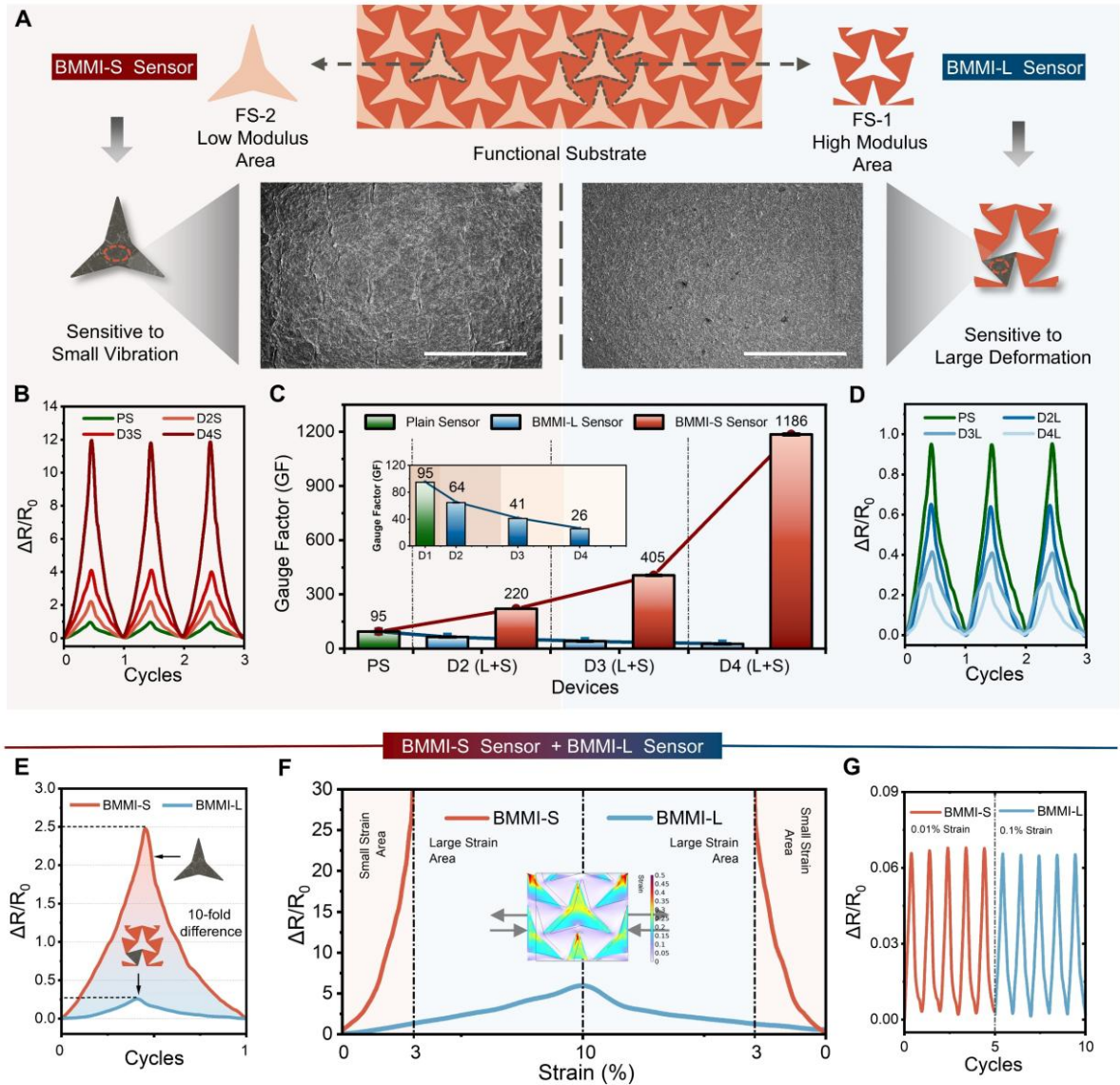

**Figure 3. Characterization of the BMMI.** (A) SEM images of the BMMI-S and BMMI-L sensors' surfaces under 5% tensile strain highlight the differentiated crack formation resulting from strain engineering. Scale bar 500μm. (B) Relative resistance changes of BMMI-S sensors in PS, D2, D3, and D4, corresponding to BMMIs with modulus ratios of 1:1, 6:1, 15:1, and 52:1, respectively, under 1% cyclic stretching. (C) Comparison of Gauge Factor (GF) between BMMI-S and BMMI-L sensors in PS, D2, D3, and D4. (D) Relative resistance changes of BMMI-L sensors in PS, D2, D3, and D4 under 1% cyclic stretching. (E) Relative resistance changes of BMMI-S and BMMI-L sensors in BMMI with a modulus ratio of 15:1, showing a tenfold difference. (F) Relative resistance changes of BMMI-S and BMMI-L sensors under 10% tensile strain and release. (G) Detection limit stability test for BMMI-S and BMMI-L sensors.

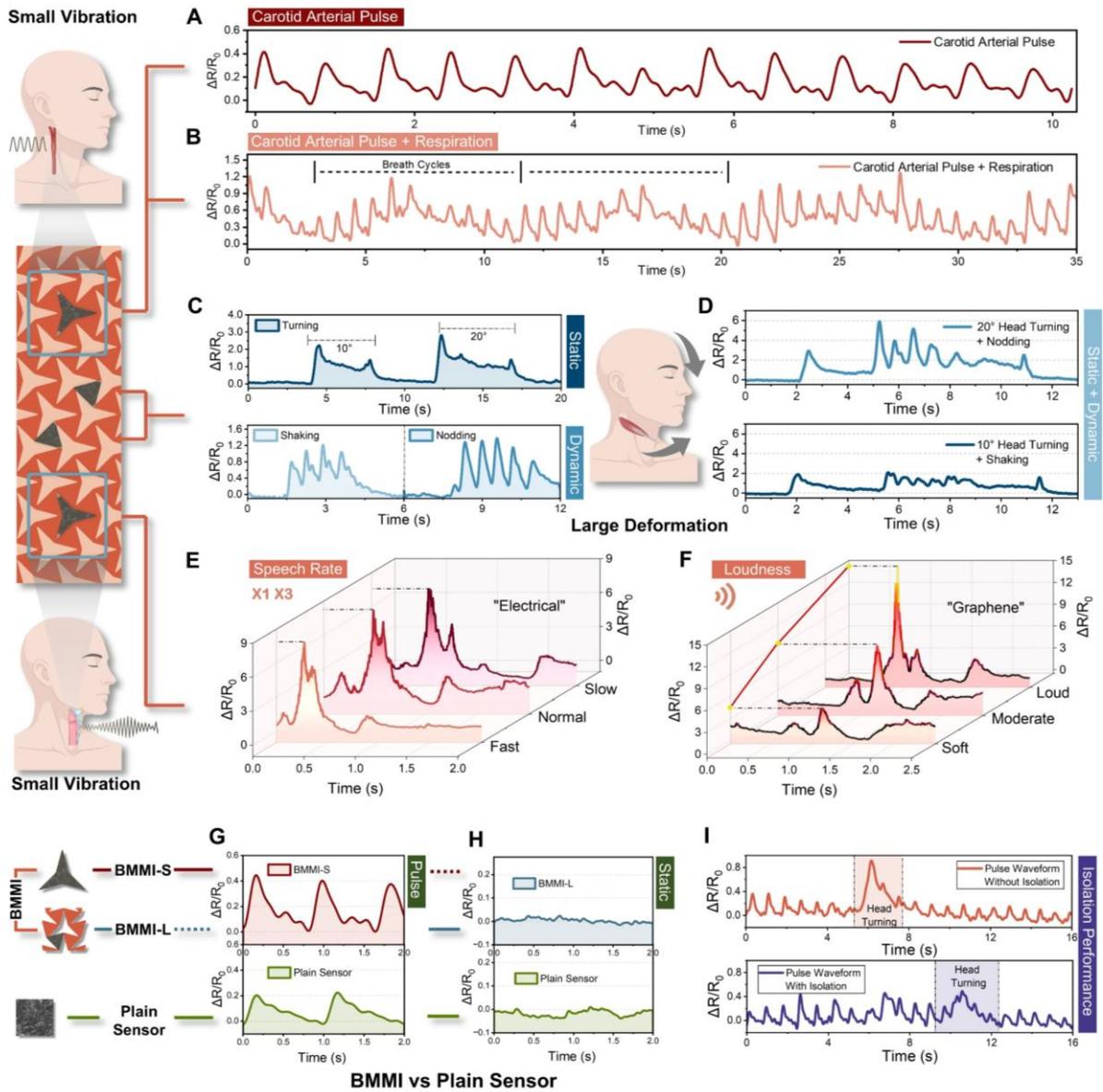

**Figure 4. Detection of Heterogeneous Mechanodermal Activities (MDAs).** (A) Carotid arterial pulse waveform captured by the BMMI-S sensor. (B) Respiratory cyclic waveform superimposed on the pulse signal captured by the BMMI-S sensor. (C) Static actions, such as turning the head 10° or 20° left, and dynamic movements like nodding and shaking, detected by the BMMI-L sensor. (D) Composite signals generated by nodding or shaking the head while it is held at a 20° or 10° leftward rotation, detected by the BMMI-L sensor. (E) Throat vibration signals of "Electrical" at fast, normal, and slow speech rates, obtained by the BMMI-S sensor. (F) Throat vibration signals of "Graphene" at soft, moderate, and loud volumes, obtained by the BMMI-S sensor. (G) Comparison of pulse signals collected by the BMMI-S and plain sensors. (H) Comparison of static signals collected by the BMMI-L and plain sensors. (I) Comparison of pulse signals during head turning, collected by the BMMI-S sensors with and without isolation frames.

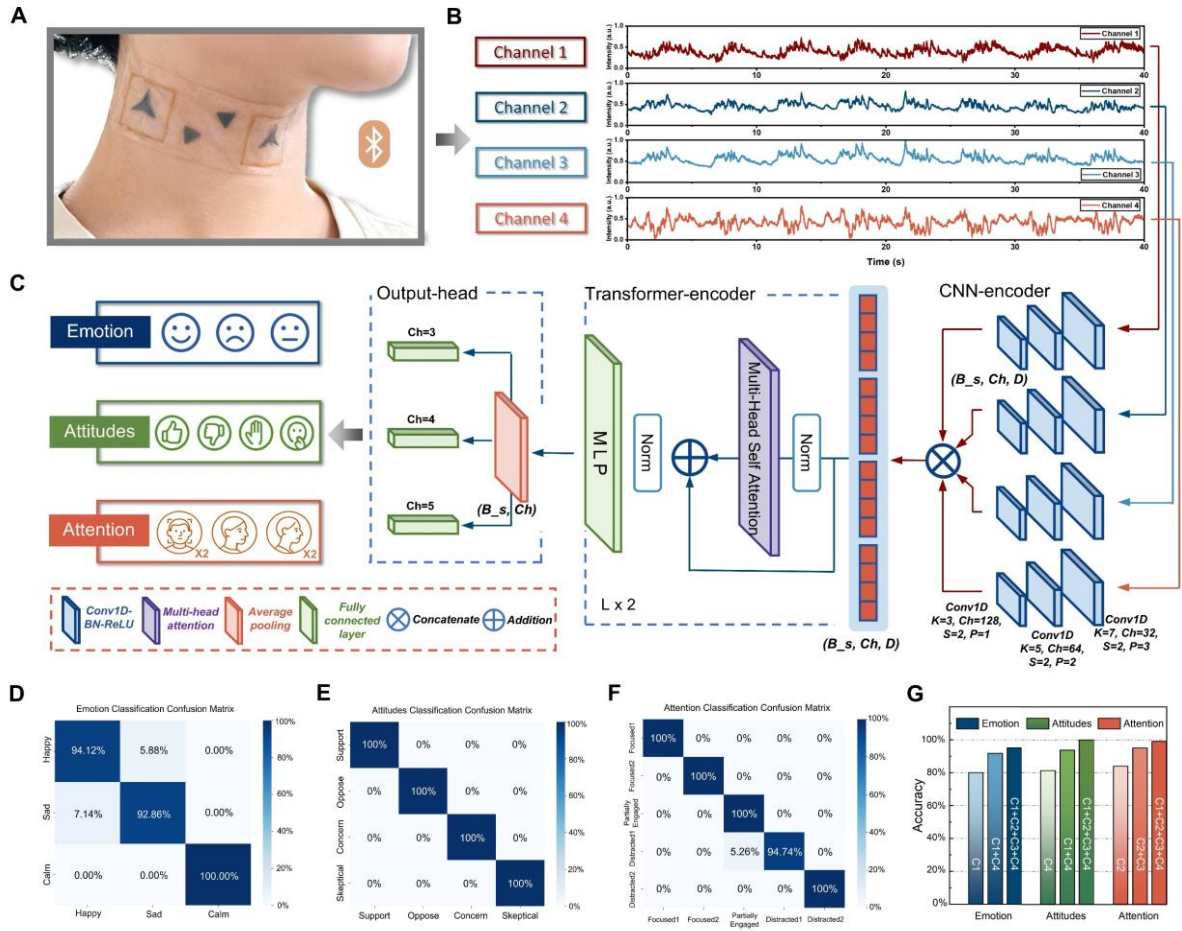

**Figure 5. Multimodal Communication Analysis by Decoding Heterogeneous MDAs.** (A) Photograph of the user wearing the four-channel wireless BMMI. (B) Four-channel signals collected by the wireless BMMI: Channel 1-subtle carotid pulse and respiratory activity; Channels 2 and 3-static and dynamic head movements; Channel 4-laryngeal vibrations. (C) The structure of the CN-Net model, based on a hybrid architecture encoder and three parallel output heads, enables precise classification of emotional, attitudinal, and attentional states. (D-F) Confusion matrices for the classification of 3 emotions, 4 attitudes, and 5 levels of attention. (G) Accuracy comparison of ablation studies conducted for the three internal states.